\documentclass[fleqn,usenatbib]{mnras}

\usepackage{newtxtext,newtxmath}
\usepackage{longtable}

\usepackage[T1]{fontenc}
\DeclareRobustCommand{\VAN}[3]{#2}
\let\VANthebibliography\thebibliography
\def\thebibliography{\DeclareRobustCommand{\VAN}[3]{##3}\VANthebibliography}

\usepackage{graphicx}	
\usepackage{amsmath}	

\title[The Stingray nebula]{Evolution of Hen\,3-1357, the Stingray Nebula\footnote{based on observations made with the Southern African Large Telescope (SALT)}}

\author[M. Pe\~na et al.]{
Miriam Pe\~na$^{1}$ \thanks{E-mail: miriam@astro.unam.mx}
Mudumba Parthasarathy$^{2,3}$,
Francisco Ruiz-Escobedo$^{1}$, and
Rajeev Manick$^{4}$
\\
$^{1}$Instituto de Astronom{\'\i}a, Universidad Nacional Aut\'onoma de M\'exico, Apdo. Postal 70254, Cd. de M\'exico, M\'exico\\
$^{2}$Indian Institute of Astrophysics, Bangalore 560034, India\\
$^{3}$Department of Physics and Astronomy, Vanderbilt University, 6301 Stevenson Center Ln., Nashville, TN 37235, USA \\
$^{4}$Univ. Grenoble Alpes, CNRS, IPAG, 38000 Grenoble, France\\
}
\date{Accepted 2022 June 21. Received 2022 June 20; in original form 2022 February 22}

\pubyear{2022}

\begin{document}
\label{firstpage}
\pagerange{\pageref{firstpage}--\pageref{lastpage}}
\maketitle

\begin{abstract}
The spectroscopic evolution of Hen\,3-1357, the Stingray Nebula, is presented by  analysing  data from 1990 to 2021. High resolution data obtained in 2021 with South African Large Telescope High Resolution Spectrograph and in 2009 with European Southern Observatory-Very Large Telescope UVES spectrograph are used to determine physical conditions and chemical abundances in the nebula. From comparison of these data with data from different epochs  it is found that the intensity of highly-ionized emission lines has been decreasing with time, while the emission of low-ionization lines has been increasing, confirming that the nebula is recombining, lowering its excitation class, as a consequence of the changes in the central star which in 2002 had an effective temperature of 60,000 K and from then it has been getting colder. The present effective temperature of the central star is about 40,000 K. It has been suggested that the central star has suffered a late thermal pulse and it is returning to the AGB phase. The nebular chemistry of Hen\,3-1357 indicates that all the elements, except He and Ne, present sub solar abundances. The comparison of the nebular abundances with the values predicted by stellar nucleosynthesis models at the end of the AGB phase, shows that the central star had an initial mass lower than 1.5 M$_{\odot}$. We estimated the ADF(O$^{+2}$) to be between 2.6 and 3.5.
\end{abstract}

\begin{keywords}
 planetary nebulae: individual: Hen 3-1357  -- stars: AGB and post-AGB -- stars: evolution
\end{keywords}



\section{Introduction}
The Stingray Nebula (also known as PN G331.3-12.1 and Hen\,3-1357) is a very young and compact planetary nebula (PN), around the rapidly evolving hot post-Asymptotic Giant Branch (post-AGB) central star SAO\,244567, whose post-AGB character was pointed out by \citet{partha89}. The compact fully-ionized nebula was discovered by \citet{partha93} and, together with the central star, they have been the subject of multiple studies due to they show unexpected variations. In particular the star, classified as a B-type star in the early 1980s, has been fading  by about 1 mag per decade \citep{schaefer15}, and also its effective temperature has shown rapid changes with time.

A relatively detailed description of the previous studies  is presented in this introduction, as it is important for our analysis.

Hubble Space Telescope (HST) high resolution images of Hen\,3-1357  showed the presence of a 1.67$\times$0.92 arcsec$^2$ nebula around the central star. The HST images also revealed the presence of collimated outflows \citep{bobrowsky98}.

A deep review of the stellar and nebular evolution can be found in \citet{balick21}. In this article the authors analysed the evolution of the nebular shape and the large decreases in the nebular emission-line fluxes based on well calibrated images obtained with HST in 1996 and 2016. They concluded that Hen\,3-1357 is now a recombination nebula. Earlier, \citet{harvey-smith18}, \citet{arkhipova13}, and \citet{otsuka17} also found evidence that the Stingray Nebula is undergoing recombination.

The central star has shown fast spectral evolution. It has evolved from a B1 type post-AGB supergiant into a PN central star in the extremely short timescale of 20 years \citep{partha93,partha95,bobrowsky98}. UV spectra of the object obtained with the International Ultraviolet Explorer (IUE)  showed the rapid evolution of the star from 1988 to 1996. P-Cygni profiles of \ion{N}{v} (1240 \AA) and \ion{C}{iv} (1550 \AA) lines in the spectra taken in 1988 indicated a terminal wind velocity of 3,500 km s$^{-1}$. According to \citet{feibelman95} and \citet{partha95} this wind seems to have completely ceased by 1994. Also these authors reported that the object faded by a factor of 3 in the UV, from 1988 to 1996. The fading suggested a rapid increase in the effective temperature, $T_{eff}$, and gravity, g, of the central star that could be rapidly evolving into a DA white dwarf \citep{partha95}. The luminosity, the core mass, the observed rapid evolution and fading of SAO\,244567 are not in agreement with the evolution timescales of low or high mass post-AGB stellar models \citep{miller16}.  

The rapid evolution of the stellar spectral type and effective temperature has been confirmed.  In 1950 the spectrum  of SAO\,244567 was that of a B or A star with weak H$\alpha$  emission \citep{henize76}. In 1971 the spectrum was that of a B1-2 supergiant star  with a $T_{eff}$ of 21,000 K and in June 1990 the spectrum was that of a planetary nebula  \citep{partha93,partha95}. From  a photoionization model of the June 1990 nebular spectrum K\"open and Parthasarathy (1996, private comm.) derived a $T_{eff}$ of 50,000 K. Thus in a matter of less than 20 years the $T_{eff}$ increased from 21,000 K to 50,000 K. 

\citet{reindl14} studied all the UV spectra of this object obtained with the IUE, since 1988 to 2006. These authors found that the central star steadily increased its $T_{eff}$ from 38,000 K to a peak value of 60,000 K in 2002, while its surface gravity increased from log(g) = 4.8 to 6.0 and there was a drop in luminosity. SAO\,244567 has cooled down significantly since 2002 and is now expanding. According to \citet{reindl14,reindl17} and \citet{lawlor21} the most reasonable explanation for the stellar variations is a late He-shell flash (late thermal pulse, LTP). The star would be now on its way back to the AGB zone.   However, respective models are lacking to match the position of SAO\,244567 in the log($T_{eff}$)–log(g) plane. The contradiction between observations and theory makes this star particularly interesting. Its fast evolution gives us the unique opportunity to study stellar evolution in real time.

\citet{harvey-smith18} analysed the full suite of Australia Telescope Compact Array data for Hen\,3-1357, taken in the 4–23 GHz range of frequencies between 1991 and 2016. The nebular radio flux density declined during that period showing signs of halting that decline between 2013 and 2016. These authors produced a spatially resolved radio image of the Stingray nebula from data obtained in 2005. A ring structure, which appears to be associated with the ring seen in HST images, was visible. In addition, they found a narrow extension of the radio emission towards the eastern and western edges of the nebula, possibly associated with a jet or outflow. The nebular  emission-measure derived by these authors  decreased between 1992 and 2011, suggesting that the nebula is undergoing recombination.

\citet{arkhipova13} carried out low-resolution spectroscopic studies of the Stingray Nebula. They present two different sets of observations, one obtained in August 1992 with the 1.5-m telescope at La Silla Observatory (European Southern Observatory, ESO-Chile) with a Boller \& Chivens spectrograph, 
and other obtained in June 2011 with the 1.9-m telescope at South African Astronomical Observatory (SAAO) and a long-slit spectrograph. 
These authors also obtained a  high-resolution spectrum that allowed them to measure a heliocentric radial velocity of $11.6 \pm 1.7$ km s$^{-1}$ and an expansion velocity of 8.4$\pm$1.5 km s$^{-1}$ for Hen\,3-1357. \citet{arkhipova13} computed the nebular physical parameters and chemical abundances concluding that Hen\,3-1357 has sub solar abundances. From the comparison of line intensities of different epochs, they found that the low-excitation lines emission (lines of O$^+$, N$^+$, and S$^+$), has increased with time, while the high-excitation lines emission (lines of O$^{+2}$, Ne$^{+2}$, and Cl$^{+2}$) decreased by factor of 2, suggesting a decrease in the excitation class of the nebula (defined as E.C. = 0.45 (F([\ion{O}{iii}]5007) / F(H$\beta$)), \citealp{dopita90}).

\citet{otsuka17} performed a detailed analysis of Hen\,3-1357 based on  high-resolution spectra obtained in 2006, covering from the optical to far-IR wavelengths. These authors calculated the nebular abundances using collisionally excited lines (CELs) and recombination lines (RLs). Their RL C/O abundance ratio would indicate that this PN is O-rich, which is also supported by the detection of the broad 9/18 $\mu$m bands from amorphous silicate grains. The observed nebular abundances can be explained by stellar AGB nucleosynthesis models with  initial masses between 1 and  1.5 M$_{\odot}$ and metallicity $\rm{Z = 0.008}$. \citet{otsuka17} reported Ne overabundance which might be due to the enhancement of $^{22}$Ne isotope in the He-rich intershell. Using the spectrum of the central star synthesized by a \textsc{tlusty} model as the ionization source of the PN, they constructed a  self-consistent photoionization model to fit the observed quantities and derive the gas and dust masses, dust-to-gas mass ratio, and core mass of the central star. According to this model about 80\% of the total dust mass is from a warm–cold dust component beyond the ionization front.  

 Since the \citet{otsuka17} study of Hen\,3-1357 in 2006 there are no high resolution spectra obtained and analysed to understand the variations in the nebula. In 2021 we obtained high resolution spectra with the South African Large Telescope High Resolution Spectrograph (SALT HRS) under the program ID:2021-1-SCI-007 (PI: Manick).  In this paper we perform a detailed analysis of these spectra and  we also analyse observations made in the ESO-Chile with the Very Large Telescope VLT-U2 and the UVES spectrograph in 2009.

 The analysis of these new data, in conjunction with studies presented by different authors along the years, allows us to discuss the evolution of the Stingray Nebula with time. This will lead to a better understanding of the evolution of the central star whose changes are affecting the nebula.

This paper is organized as follows: In \S2 the new observations and data reduction are presented. In \S3 we discuss the time evolution of line intensities. In \S4 the physical parameters and the ionic abundances of the nebula are determined from the intensities of our observed CELs and RLs, which will allow to determine the Abundance Discrepancy Factors (ADF)\footnote{The ADF is defined as the ratio between abundances derived from RLs and abundances derived from CELs.}.  Also the time evolution of the physical parameters and ionic abundances is presented in this section.  Total nebular abundances derived from our work are presented in \S5 and they are compared with abundances from other authors. In \S6 the discussion can be found. In \S7 our conclusions are presented.

\section{Observations and data reduction}

As said above, the nebula Hen\,3-1357 has been observed and studied in multiple occasions. 
 In this work we analyse spectrophotometric data obtained with ESO VLT UVES and SALT HRS spectrographs, performed in the years 2009 and 2021, respectively.
 
 ESO VLT UVES is a high-resolution optical echelle spectrograph located at the Nasmyth B focus of telescope UT2. The light beam of the telescope is split in two arms (blue and red). The resolving power is about 40,000 when a slit width of 1 arcsec is used.  The  data analysed here were retrieved from the archives, already wavelength and flux calibrated with the procedures described at ESO Data Release 2020. The slit size for the observations was 1$\times$10 arcsec and the extraction slit of the spectra was 39 pix, equivalent to 2.3 arcsec, which implies that all the nebular emission in the slit was extracted.

SALT HRS is an echelle spectrograph working with two arms simultaneously. The blue one covers from 3826 \AA~ to 5583 \AA~ wavelength range and the red one covers from 5473 \AA~ to 8796 \AA \, range. Slit size used for the nebula  was 1.2 arcsec (fiber diameter 350 $\mu$m). The spectral resolution  R (defined as $\lambda / \delta\lambda$) was about 65,000 at 6,000 \AA. Three spectra  were obtained in May-June 2021. The observations were made with the slit in parallactic angle to minimize the influence of atmospheric dispersion. 
 
Both observing logs, ESO VLT-U2-UVES and SALT HRS, are presented in Table \ref{tab:log_table}.
 
SALT data were initially calibrated with the HRS pipeline described by \citet{kniazev16} which includes bias subtraction, flat-fielding, extraction, wavelength calibration and merging of different orders. Also heliocentric velocity correction was calculated for each spectrum. Afterwards, data were corrected by atmospheric absorption using SALT absorption law and were flux-calibrated by using \textsc{iraf}\footnote{IRAF is distributed by the National Optical Astronomy Observatories, which is operated the Association of Universities for Research in Astronomy, Inc., under contract to the National Science Foundation.} standard routines. The flux standard star BD+02d3375, observed with SALT HRS on 2021 July 18th, was used for flux calibration by adopting the calibrated stellar fluxes obtained from HST-STIS CALSPEC Calibration Database as suggested by  \citet{kniazev17}.  By employing this standard star a relative spectral calibration can be obtained. No absolute flux calibration is possible since SALT is a telescope with a variable aperture.
 
\begin{figure}
	\includegraphics[width=\columnwidth]{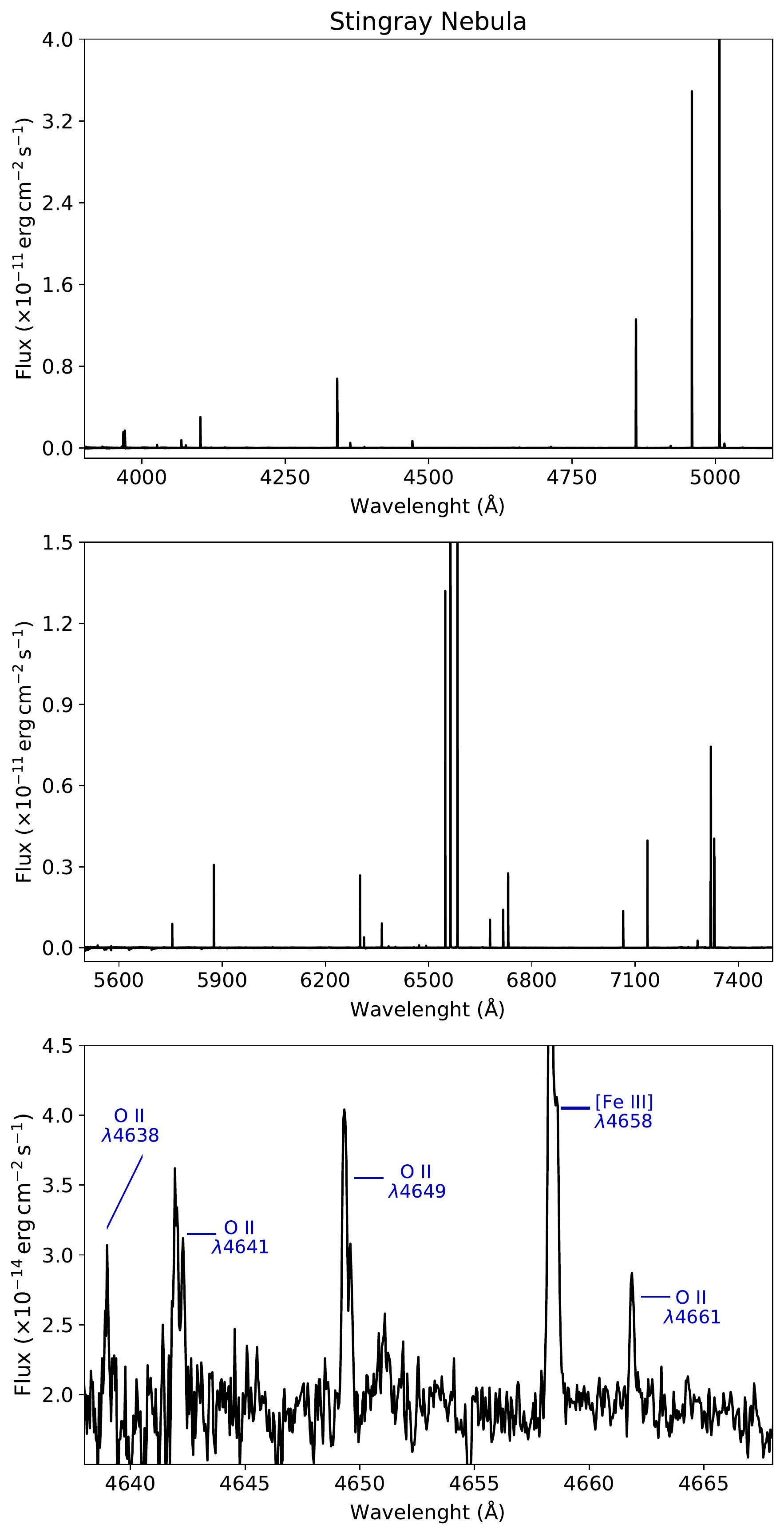}
    \caption{SALT blue and red spectra of Stingray nebula, obtained in 2021. Bottom:  SALT spectrum showing the faint recombination lines in the zone 4640 \AA~ to 4670 \AA. }
    \label{fig:spectra}
\end{figure}

The calibrated SALT HRS spectra were combined in one, to improve the signal-to-noise.
The blue and red zones of the combined spectra are shown in Fig. \ref{fig:spectra}. The bottom figure shows the faint recombination lines in the 4640 \AA~ to 4690 \AA~ zone, where the most intense recombination lines of \ion{O}{ii} are located.  The intensity of these lines is thousandths or ten thousandths times lower than H$\beta$ and it is clear that they are perfectly detected  and measurable, therefore this figure shows the high quality of the SALT HRS spectra.

The line intensities in ESO VLT UVES and SALT HRS spectra were measured with the \textsc{iraf}'s \textit{splot} routine, by integrating the flux between two given limits over a continuum estimated by eye. Due to the high spectral resolution most lines appear isolated, no blends are affecting our line measurements. More than a hundred nebular lines were measured in the wavelength range from 3826 \AA~ to 8796 \AA. Dereddened fluxes from 3770 \AA~ to 7751 \AA~ are presented in columns 6 and 8 of Table \ref{tab:intensities}. Uncertainties of intensities were determined taking into account the SNR of the continuum at each side of the line and the line intensity. The uncertainties  are included in parenthesis in each case. In the other columns we present the line intensities given by other authors in different epochs.
 
The logarithmic reddening correction, c(H$\beta$),  was derived from our data by using \citet{cardelli:89} reddening law by assuming a ratio of total to selective extinction $\mathrm{R_{V} = 3.1}$ and by using the H Balmer decrement (theoretical H$\gamma$/H$\beta$ and H$\delta$/H$\beta$ ratios) given by \citet{storey95} for a temperature of 10,000 K, adequate to the value derived for Hen\,3-1357. In all cases, case B recombination theory and a density of $10^3$ cm$^{-3}$ were assumed. Values for c(H$\beta$) derived in this work, and those presented by other authors, are included in Table \ref{tab:intensities}.
 
 \section{Time evolution of line fluxes}
 
Dereddened fluxes at each wavelength, for the different observations from 1992 to 2021 are listed in Table \ref{tab:intensities}, as they were presented by the different authors (no uncertainties have been included for other authors). 
Along the time important changes in the line intensities are found in this table. It is evident that the collisionally excited [\ion{O}{iii}] lines 5007, 4959 and 4363 \AA~ have decreased  by a factor larger than 2 from 1992 to 2021. 
Also [\ion{Ne}{iii}] 3869 and 3967 \AA~ have decreased by factors of 2.9. Additionally
the highly ionized lines of [\ion{Ar}{iv}] 4711 and 4740 \AA~ have disappeared and they are not detected in 2021 data. Lines of [\ion{Cl}{iii}] 5517 and 5538 \AA~ decreased by about 1.8 or more and [\ion{S}{iii}] 6312 \AA~ has decreased by 22\%.
In addition, \ion{He}{i} recombination  lines have also diminished.
\ion{He}{i} 5876 \AA~ decreased by more than 30\%, while \ion{He}{i} 3867 \AA~ decreased by a factor of almost 2. 

From the above it is concluded that the intensity of highly-excited lines, listed in the paragraph above, have been decreasing systematically.
On the contrary, line intensities from low-ionized species have increased. That is the case of  [\ion{Fe}{iii}] 4755 and 4881 \AA, [\ion{N}{ii}] 6548 and 6583 \AA, and [\ion{S}{ii}] 6716 and 6730 \AA~ whose intensities increased by factors of about 2. Other lines that increased substantially are the [\ion{O}{i}] 6300 and 6363 \AA~ lines which in principle do not belong to the ionized nebula and are possibly emitted in an external photo-dissociation zone.

\begin{figure}
	\includegraphics[width=\columnwidth]{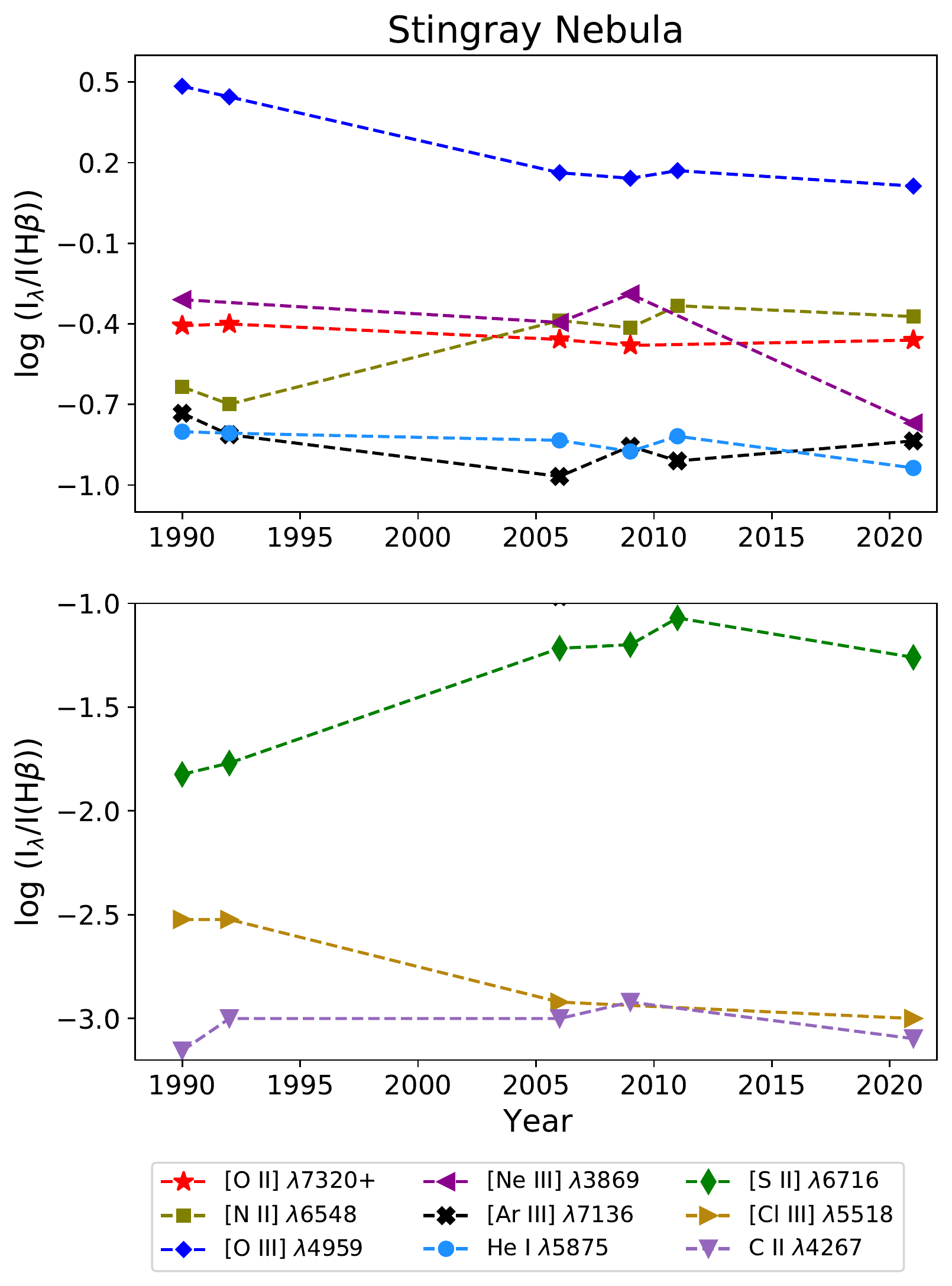}
    \caption{The time evolution of some important lines intensities are plotted. The lines are identified by colours. The upper panel shows the more intense lines.
	}
    \label{fig:line-evolution}
\end{figure}

In Fig. \ref{fig:line-evolution} the time evolution of some important lines are shown.
 These line intensity variations are indicating that the nebular excitation class has diminished from about 9 in 1990-1992 to about 4 in 2021. The highly-ionized species are recombining, increasing the ionic abundances of low-ionized species. This evolution is a reaction of the plasma to the fact that the central star is getting colder, probably in its way returning to the AGB zone as a result of a late thermal pulse, LTP, as suggested by \citet{reindl14,reindl17} and \citet{lawlor21}.

\subsection{RL evolution}
\ion{He}{i} lines kept the same intensities up to about 2011. In the recent observations with SALT HRS  in 2021, these intensities have decreased, most probably due to recombination of He$^+$ into He$^0$. This is confirmed by the fact that He/H total abundance appears lower on 2021, due to only the He$^+$/H$^+$ abundance ratio is considered and  neutral He is not taken into account (see Table \ref{tab:chemistry}).

The \ion{C}{ii} 4267 \AA \, recombination line, employed to determine C$^{+2}$/H$^+$ ionic abundance, and the recombination lines of \ion{O}{ii}, useful to determine  O$^{+2}$/H$^+$ ionic abundance, have remained constant.

\begin{table*}
	\centering
	\caption{Log of ESO-VLT-U2-UVES and SALT HRS observations of Hen\,3-1357}
	\label{tab:log_table}
	\begin{tabular}{lccccrrr} 
		\hline
		Spectrum & Obs. date & Wavelength range (\AA)& Spectral res. & Slit width ('') & Exp.time (s) & Air mass & Seeing ('')\\
		\hline
		ESO-VLT-U2-UVES\\
		ADP.2020-06-26T08-14-43.698& 2009-03-16& 3730-5000 & 40,970 & 1.0 &600 & 1.363& 1.74  \\
		ADP.2020-06-26T08-14-43.724& 2009-03-16& 5650-9450 & 42,310 & 1.0 & 600 & ''&  1.74  \\
		\hline
		SALT-HRS \\
		mbgphH202105240022 & 2021-05-24 & 3826 - 5583 & 65,000 & 1.2 & 1200 & 1.170 & 1.8\\
		mbgphR202105240022 & 2021-05-24 & 5473 - 8796  &    ''& ''& ''  & ''& 1.8\\
		mbgphH202105250025 & 2021-05-26 & 3826 - 5583 &   ''& '' & 600 & 1.203 & 1.7 \\
		mbgphR202105250025 & 2021-05-26 & 5473 - 8796  &  '' & '' &   '' & ''& 1.7\\
		mbgphH202106240100 & 2021-06-24 &  3826 - 5583 & ''& '' & 600 & 1.103& 1.7\\
		mbgphH202106240100 & 2021-06-24 & 5473 - 8796  & '' &  '' &  '' &   ''& 1.7\\
		\hline
	\end{tabular}
\end{table*}

\begin{table*}
\centering
\label{tab:intensities}
\caption{Line intensities in different epochs, as presented by the different authors$^{a,b}$}
\begin{tabular}{rcrrrrrr}
\hline 
   & & Parthasarathy & Arkhipova   & Otsuka  et al. & ESO UVES & Arkhipova   & SALT HRS \\
  & & et al. (1993) & et al. (2013) & (2017)   & this work &  et al. (2013) & this work  \\
  \hline
 & c(H$\beta$) & 0.26 & 0.19 & 0.08 & 0.15 & 0.35 & 0.22 \\
\hline
 ion & $\lambda_0$ & obs 1990 & obs 1992 & obs 2006 & obs 2009 & obs 2011 & obs 2021 \\
 \hline
~[\ion{O}{ii}] & 3727.00 & 81.20 & 58.10 & 138.86 & --- & 163.00 & --- \\
H11 & 3770.63 & --- & --- & 4.35 & 4.86(0.30) & --- & --- \\ 
H10 & 3797.90 & 4.20 & 4.70 & 5.33 & 6.13(0.30) & --- & --- \\ 
\ion{He}{i} & 3819.60 & --- & --- & 1.14 & 1.40(0.20) & --- & --- \\ 
\ion{He}{i} & 3833.35 & --- & --- & 0.07 & 0.07(0.02) & --- & --- \\ 
H9 & 3835.38 & 6.00 & 6.40 & 7.59 & 7.94(0.40) & --- & 6.00(1.20)   \\ 
\ion{He}{i} & 3867.47 & --- & --- & 0.15 & 0.18(0,07) & --- & 0.08(0.04)   \\ 
~[\ion{Ne}{iii}] & 3869.06 & 49.00 & --- & 40.30 & 51.40(1.02) & --- & 17.01(1.00)   \\ 
\ion{He}{i} & 3871.79 & --- & --- & 0.08 & 0.09(0.01) & --- & ---   \\ 
H8 & 3889.05 & 17.20+ & 18.9+ & 16.00 & 20.36(1.00) & 9.20 & 20.62(1.20)   \\ 
\ion{He}{i} & 3926.54 & --- & --- & 0.12 & 0.13(0.02) & --- & 0.18(0.10)   \\ 
\ion{He}{i} & 3964.73 & --- & --- & 0.54 & 0.94(0.05) & --- & 0.90(0.09)   \\ 
~[\ion{Ne}{iii}] & 3967.79 & --- & --- & 10.00 & 15.45(0.90) & --- & 8.85(0.08)   \\ 
H7 & 3970.07 & 32.30+ & 40.1+ & 16.03 & 16.23(0.90) & --- & 15.71(0.07)   \\ 
\ion{He}{i} & 4009.26 & --- & --- & 0.15 & 0.20(0.05) & --- & ---   \\ 
\ion{He}{i} & 4026.20 & 1.90 & 2.10 & 1.53 & 2.38(0.05) & --- & 2.42(0.12)   \\ 
~[\ion{S}{ii}] & 4068.60 & 3.20 & 3.50 & 4.47 & 6.43(0.90) & --- & 4.93((0.25)   \\ 
\ion{O}{ii} & 4069.62 & --- & --- & 0.27 & 0.08(0.06) & --- & 1.67(0.30)   \\ 
\ion{O}{ii} & 4069.88 & --- & --- & 0.38 & 0.06(0.04) & --- & ---   \\ 
\ion{O}{ii} & 4075.86 & --- & --- & 0.76 & 0.09(0.05) & --- & ---   \\ 
~[\ion{S}{ii}] & 4076.35 & 0.60 & 0.60 & 1.51 & 2.28(0.44) & --- & ---   \\ 
\ion{N}{iii} & 4097.35 & --- & --- & 0.02 & 0.03(0.02) & --- & ---   \\ 
H$\delta$ & 4101.73 & 24.50 & 26.70 & 21.52 & 25.55(0.90) & 21.60 & 26.11(0.50)   \\ 
\ion{O}{ii} & 4119.22 & --- & --- & 0.02 & 0.02(0.01) & --- & ---   \\ 
\ion{He}{i} & 4120.81 & --- & --- & 0.18 & 0.23(0.03) & --- & ---   \\ 
\ion{He}{i} & 4143.76 & --- & --- & 0.23 & 0.34(0.03) & --- & 0.14   \\ 
\ion{O}{ii} & 4153.30 & --- & --- & 0.03 & 0.03(0.01) & --- & ---   \\ 
\ion{C}{ii} & 4267.18 & 0.07 & 0.10 & 0.10 & 0.12(0.05) & --- & 0.08(0.04)   \\ 
H$\gamma$ & 4340.46 & 43.20 & 45.70 & 46.05 & 46.67(1.00) & 50.90 & 47.34(0.50)   \\ 
\ion{O}{ii} & 4349.43 & --- & --- & 0.03 & 0.05(0.02) & --- & 0.09(0.06)   \\ 
~[\ion{O}{iii}] & 4363.21 & 7.50 & 7.90 & 2.46 & 3.26(0.25) & 4.60 & 2.73(0.35)   \\ 
\ion{He}{i} & 4387.93 & --- & --- & 0.44 & 0.65(0.07) & --- & 0.70(0.40)   \\ 
\ion{He}{i} & 4437.55 & --- & --- & 0.07 & 0.08(0.04) & --- & ---   \\ 
\ion{He}{i} & 4471.47 & --- & 5.50 & 4.63 & 5.23(0.40) & --- & 5.55(0.50)   \\
\ion{N}{ii} & 4630.54 & --- & --- & 0.02 & 0.02(0.01) & --- & ---   \\ 
\ion{O}{ii} & 4638.86 & --- & --- & 0.03 & 0.04(0.02) & --- & 0.04(0.03)   \\ 
\ion{O}{ii} & 4641.81 & --- & --- & 0.06 & 0.08(0.04) & --- & 0.12(0.02)   \\ 
\ion{O}{ii} & 4649.13 & --- & --- & 0.10 & 0.13(0.07) & --- & 0.12(0.02)   \\ 
\ion{O}{ii} & 4650.84 & --- & --- & 0.03 & 0.03(0.02) & --- & ---   \\ 
~[\ion{Fe}{iii}] & 4658.05 & --- & --- & 0.11 & 0.17(0.08) & --- & 0.29(0.03)   \\ 
\ion{O}{ii} & 4661.63 & --- & --- & 0.05 & 0.04(0.02) & --- & 0.04(0.02)   \\ 
\ion{O}{ii} & 4676.23 & --- & --- & 0.03 & 0.03(0.02) & --- & 0.05(0.03)   \\ 
~[\ion{Fe}{iii}] & 4701.53 & --- & --- & 0.05 & 0.06(0.03) & --- & 0.10(0.05)   \\ 
~[\ion{Ar}{iv}] & 4711.37 & --- & --- & 0.03 & 0.04(0.01) & --- & < 0.01   \\ 
\ion{He}{i} & 4713.22 & 0.90 & 1.00 & 0.67 & 0.79(0.08) & --- & 0.85(0.09)   \\ 
~[\ion{Fe}{iii}] & 4733.91 & --- & --- & 0.03 & 0.02(0.01) & --- & ---   \\ 
~[\ion{Ar}{iv}] & 4740.16 & --- & --- & 0.07 & 0.06(0.03) & --- & < 0.01   \\ 
~[\ion{Fe}{iii}] & 4754.69 & --- & --- & 0.03 & 0.03(0.01) & --- & 0.06   \\ 
H$\beta$ & 4861.33 & 100.00 & 100.00 & 100.00 & 100.00(1.00) & 100.00 & 100.00(1.00)   \\
~[\ion{Fe}{iii}] & 4881.00 & --- & --- & 0.05 & 0.08(0.02) & --- & 0.15(0.07)   \\ 
\ion{He}{i} & 4921.93 & --- & --- & 1.21 & 1.45(0.15) & --- & 1.45(0.12)   \\ 
~[\ion{O}{iii}] & 4931.23 & --- & --- & 0.06 & 0.06(0.03) & --- & 0.10(0.05)   \\ 
~[\ion{O}{iii}] & 4958.91 &305.00 &278.50 &145.40 &138.59(1.00) &148.00 & 129.65(1.30) \\ 
~[\ion{O}{iii}] & 5007.06 & 933.00 & 860.30 & sat & oor & 443.90 & 411.91(2.00)   \\ 
~[\ion{Ar}{iii}] & 5191.82 & 0.60 & --- & 0.06 & oor & --- & 0.09(0.04)   \\ 
~[\ion{Cl}{iii}]  & 5517.72 & 0.30 & 0.30 & 0.12 & oor & --- & 0.10(0.03)   \\
~[\ion{Cl}{iii}]  & 5537.89 & 0.40 & 0.40 & 0.24 & oor & --- & 0.22(0.02)   \\
\hline
\multicolumn{6}{l}{$^a$ In the Table, 'sat' indicates saturated line, `oor': out of range, not observed.}\\
\multicolumn{6}{l}{$^b$ A '+' indicates blend of two lines.} 
\end{tabular}
\end{table*}

\begin{table*}
\centering
\contcaption{}
\begin{tabular}{rcrrrrrr}
\hline
    & & Parthasarathy & Arkhipova   & Otsuka  et al. & ESO UVES & Arkhipova  & SALT HRS \\
  & & et al. (1993) & et al.(2013) & (2017)   & this work & et al. (2013) & this work  \\
\hline
 ion & $\lambda_0$ & obs 1990 & obs 1992 & obs 2006 & obs 2009 & obs 2011 & obs 2021 \\
 \hline
~[\ion{N}{ii}] & 5754.64 & 1.80 & 2.40 & 2.58 & 3.15(0.15) & 3.30 & 2.72(0.12)   \\ 
\ion{He}{i} & 5875.60 & 15.80 & 15.60 & 14.67 & 13.34(0.90) & 15.20 & 11.58(0.55) \\ 
~[\ion{O}{i}] & 6300.30 & 4.10 & 7.40 & 16.13 & 14.59(0.90) & 17.70 & 12.28(0.55)   \\ 
~[\ion{S}{iii}] & 6312.10 & 1.40 & 1.70 & 1.03 & 1.28(0.15) & 1.20 & 1.14(0.08)   \\ 
~[\ion{O}{i}] & 6363.78 & 2.00 & 2.40 & 5.10 & 5.46(0.30) & 6.50 & 4.18(0.12)   \\ 
~[\ion{N}{ii}] & 6527.24 & --- & --- & 0.02 & 0.03(0.02) & --- & ---   \\ 
~[\ion{N}{ii}] & 6548.04 & 23.20 & 20.00 & 40.96 & 38.52(1.00) & 46.50 & 42.39(0.84)   \\ 
H$\alpha$ & 6562.82 & 305.00 & 286.20 & sat & sat & 285.50 & 286.75(2.50)   \\ \ion{C}{ii} & 6578.05 & --- & --- & 0.05 & 0.06(0.03) & --- & 0.07(0.05)   \\ 
~[\ion{N}{ii}] & 6583.46 & 78.60 & 64.90 & 121.45 & sat & 144.60 & 135.22(2.60)   \\ 
\ion{He}{i} & 6678.15 & 4.40 & 3.80 & 3.97 & 4.54(0.30) & 4.20 & 3.74(0.25)   \\ 
~[\ion{S}{ii}] & 6716.44 & 1.50 & 1.70 & 6.07 & 6.32(0.40) & 8.50 & 5.49(0.20)   \\ 
~[\ion{S}{ii}] & 6730.81 & 3.00 & 3.40 & 12.47 & 12.20(0.90) & 16.10 & 10.56(0.30)   \\ 
\ion{O}{i} & 7002.12 & --- & --- & 0.05 & 0.05(0.02) & --- & 0.06(0.04)   \\ 
\ion{He}{i} & 7062.28 & --- & --- & 0.02 & 0.02(0.01) & 0.00 & 0.02(0.02)   \\
\ion{He}{i} & 7065.18 & 9.00 & 8.50 & 7.85 & 8.46(0.40) & 7.40 & 6.27(0.10)   \\ 
~[\ion{Ar}{iii}] & 7135.78 & 18.50 & 15.40 & 10.77 & 13.95(0.80) & 12.30 & 14.60(0.70) \\ 
~[\ion{Fe}{ii}] & 7155.16 & --- & --- & 0.03 & 0.05(0.02) & --- & 0.05(0.01)   \\ 
~[\ion{He}{i}] & 7160.61 & --- & --- & 0.03 & 0.03(0.01) & --- & 0.04(0.01)  \\ 
\ion{He}{i} & 7281.35 & --- & 0.80 & 0.73 & 0.93(0.05) & --- & 1.19(0.60)   \\
~[\ion{O}{ii}] & 7318.92 & --- & --- & 4.00 & 5.61(0.30) & --- & 6.29(0.40)   \\ 
~[\ion{O}{ii}] & 7319.99 & 21.8+ & 18.0+ & 13.49 & 17.73(1.00) & --- & 18.99(0.70)   \\ 
~[\ion{O}{ii}] & 7329.66 & ---   & ---   & 7.48  & 10.34(1.00) & --- & 10.64(0.70)  \\ 
~[\ion{O}{ii}] & 7330.73 & 17.4+ & 14.3+ & 6.99 & 9.76(1.30) & --- & 9.36(0.40)   \\ 
~[\ion{Ni}{ii}] & 7377.83 & --- & --- &   0.03 & 0.03(0.01) & --- & 0.03(0.01)   \\ 
\ion{He}{i} & 7499.85 & --- & --- & 0.04   &  0.04(0.01) & --- & 0.04(0.01)   \\ 
~[\ion{Ar}{iii}] & 7751.10 & --- & 3.60 & 2.63   & & --- & 2.99(0.15)   \\ 
\hline
\end{tabular}
\end{table*}
 
 \section{Physical parameters of the nebula}

From the nebular lines, in particular from those collisionally excited lines, physical conditions, such as electron densities and temperatures, can be derived from some diagnostic line ratios.  In general electron densities can be determined from the [\ion{S}{ii}] $\lambda\lambda$6731/6716, [\ion{O}{ii}] $\lambda\lambda$3729/3726, [\ion{Cl}{iii}] $\lambda\lambda$5538/5518,  [\ion{Fe}{iii}] $\lambda\lambda$4701/4659, and [\ion{Ar}{iv}] $\lambda\lambda$4711/4740 intensity ratios. Electron temperatures can be obtained from the [\ion{N}{ii}] $\lambda\lambda$(6548+6583)/5755, [\ion{O}{iii}] $\lambda\lambda$(5007+4959)/4363, [\ion{Ar}{iii}] $\lambda\lambda$7136/5192,  [\ion{Ar}{iv}] $\lambda\lambda$(7170+7263)/(4711+4740), [\ion{S}{ii}] $\lambda\lambda$(6716+6731)/(4068+4076) and [\ion{O}{ii}] $\lambda\lambda$7325/3727 intensity ratios. The intensity of the auroral line [\ion{N}{ii}] $\lambda$5755 was corrected by effects of recombination of N$^{+2}$ using the procedure presented by \citet{liu00}, by using the ORL abundance of N$^{+2}$ and the temperature determined for CELs, both presented in Table \ref{tab:chemistry}; this could be applied only to UVES-2009 data, in which the contribution of recombination to [\ion{N}{ii}] $\lambda 5755$ line intensity is about 1\%. 

From the available diagnostic line ratios of the observations presented here, physical conditions were calculated with the code \textsc{PyNeb} \citep{luridiana15}, using the atomic data presented in the Appendix B. \textsc{PyNeb} routine \textit{getCrossTemden} was used to determine simultaneously temperatures and densities by building diagnostic diagrams.  The uncertainties in the physical conditions and abundances were estimated using Monte Carlo simulations using 400 random points, 
assuming a normal distribution around each line intensity bounded by the observed flux error.

\subsection{Diagnostics diagrams}
\begin{figure}
	\includegraphics[width=\columnwidth]{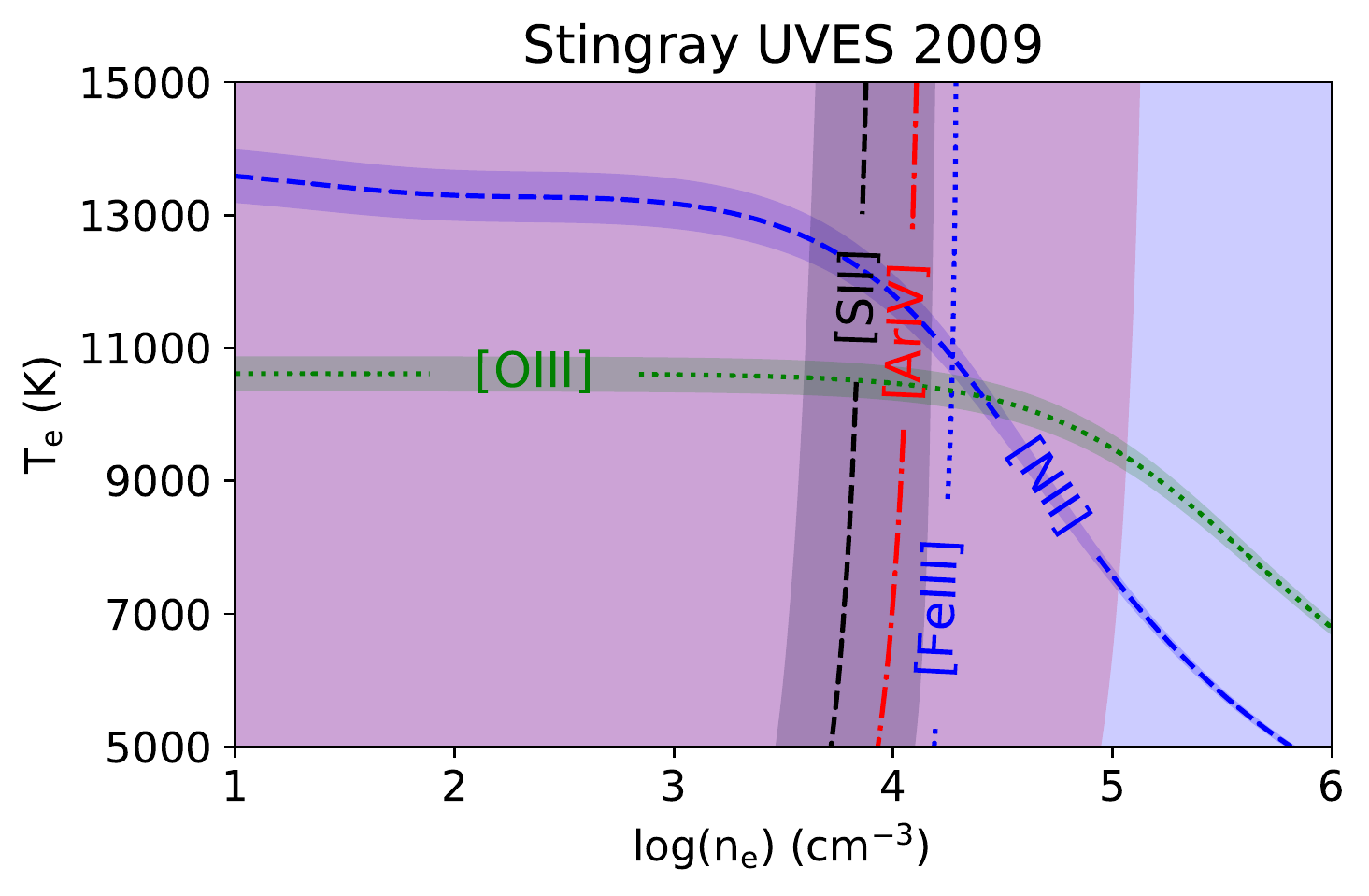}
	\includegraphics[width=\columnwidth]{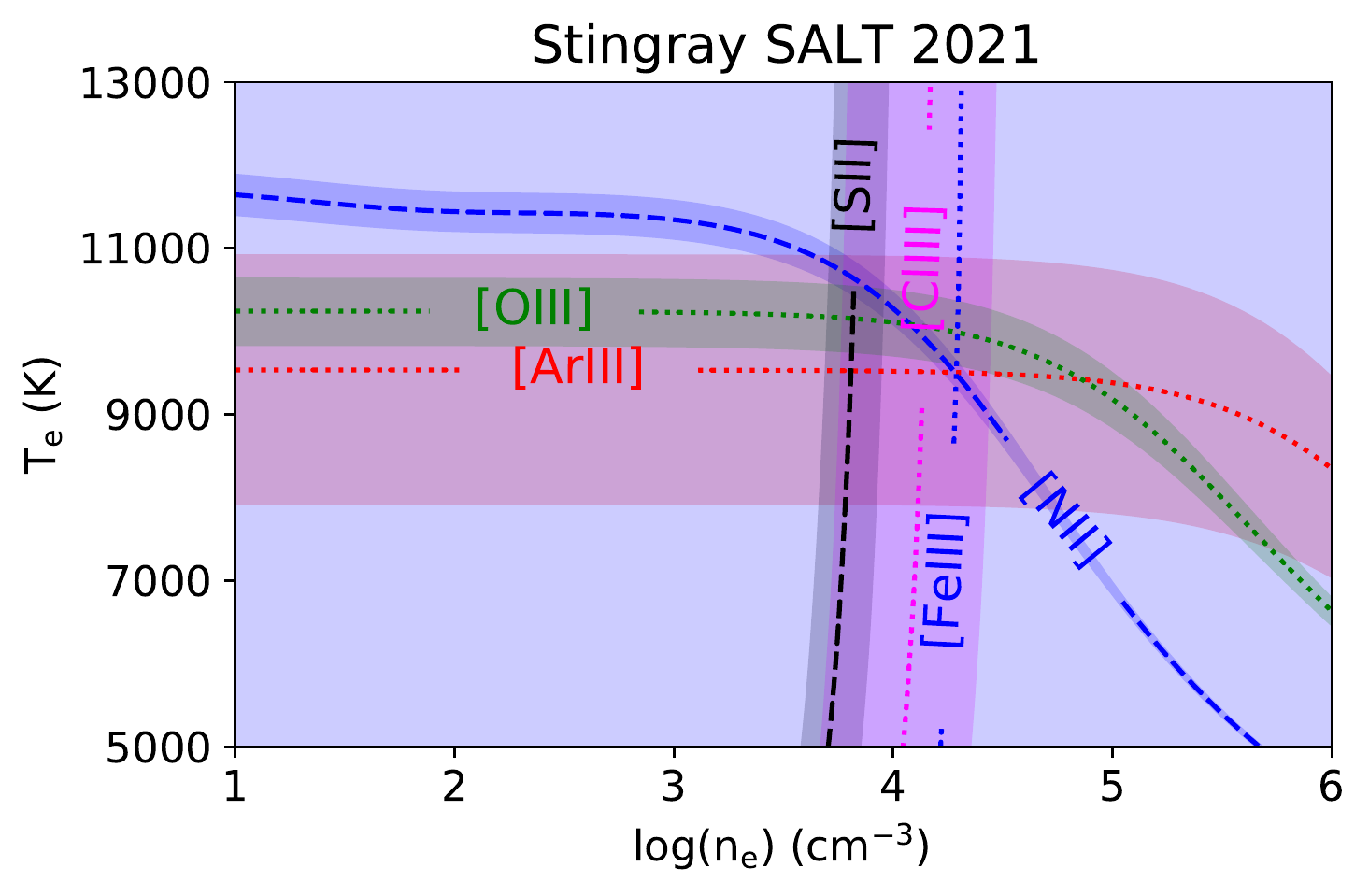}
    \caption{The diagnostic diagrams for ESO VLT-UVES and SALT HRS data, constructed by using \textsc{PyNeb}, are presented. The diagnostic lines ratios are plotted as a function of the electron temperature and density. They are represented as broken or dotted lines inside a colour band showing the $1 \sigma$ rms error. 
	}
    \label{fig:diagnostics}
\end{figure}

In Fig. \ref{fig:diagnostics} diagnostic diagrams for the  ESO VLT-UVES and SALT HRS data analysed here are presented.  These diagrams were constructed by using PyNeb, and they show the behaviour of diagnostic line ratios, such as [\ion{O}{iii}]$\lambda\lambda$(4959+5007)/4363 and [\ion{S}{ii}]$\lambda\lambda$6731/6716,  as a function of the electron density and temperature. 

From the ESO VLT UVES data the available line ratios for temperature determination were those from [\ion{N}{ii}]$\lambda\lambda$(6548+6584)/5755, and [\ion{O}{iii}]$\lambda\lambda$(4959+5007)/4363 and the density could be derived from [\ion{S}{ii}]$\lambda\lambda$6731/6716, [\ion{Ar}{iv}]$\lambda\lambda$4711/4740  and [\ion{Fe}{iii}]$\lambda\lambda$4701/4659. In the case of SALT HRS data, the temperature was determined from [\ion{N}{ii}], [\ion{O}{iii}] and [\ion{Ar}{iii}] diagnostic line ratios, while the density was derived from [\ion{S}{ii}], [\ion{Cl}{iii}] and [\ion{Fe}{iii}] diagnostic line ratios. 

Each line ratio in the diagram is represented by a broken or dotted line inside a colour band  which shows the 1$\sigma$ rms error.   The  temperature and density values derived from our observations are listed in Table \ref{tab:chemistry}. 
 Usually the electron densities and temperatures adopted for ionic abundance determination are obtained from the zone where density- and temperature-diagnostics intersect. In the diagnostic diagrams it is observed that densities from [\ion{Ar}{iv}] and [\ion{Fe}{iii}] are very large and uncertain and will not be used for this purpose.

\defcitealias{kingsburgh94}{KB94}
\defcitealias{delgado-inglada14}{DI14}
\defcitealias{peimbert14}{P14}
\defcitealias{rodriguez05}{RR05}

\begin{table*}
\caption{ Physical conditions, ionic and total abundances}
\label{tab:chemistry}
\begin{tabular}{lrrrrrr}
\hline
\hline 
Authors & Partha. 1993 & Arkhipova  &  Otsuka & ESO-UVES & Arkhipova & SALT HRS \\
Obs. year  & 1990 & 1992 & 2006 & 2009 & 2011 & 2021 \\
\hline
$T_e$([\ion{N}{ii}]) & 11,000(250) & 12,938(2,444) &  9,280(100) & 12,200$_{-800}^{+600}$ & 11,066(1752) & 10,200$_{-600}^{+400}$ \\
$T_e$([\ion{O}{iii}]) & 11,000(250) & 11,051(755) & 9,420(40) & 10,500$_{-300}^{+200}$ &  11,553(1579) & 9,900$\pm$400 \\
$T_e$([\ion{Ar}{iii}]) &  & --- & 8,670(150) & --- &  --- & 9,600$_{-1,400}^{+1,300}$ \\
$T_e$(\textit{mean}) & ---  & --- &  --- &11,300$_{-500}^{+300}$  & --- & 9,900$_{-600}^{+500}$ \\
\hline 
$n_e$ CELs &  &  &  &  &  & \\
$n_e$([\ion{S}{ii}]) & 10,000 & 15,796(10,398) & 5,710(1790) & 6,800$_{-2800}^{+6500}$ &  8,740(7701) & 6,500$_{-1,700}^{+2,600}$ \\
$n_e$([\ion{Cl}{iii}]) &  & 7416(5865) & 23,970(3120) & --- &  --- & 13,600$_{-6,400}^{+18,800}$ \\
$n_e$([\ion{Ar}{iv}]) & ---  & --- & 22 720(4360) & 31,400$_{-23,300}^{+102,900}$ &  --- & --- \\
$n_e$([\ion{Fe}{iii}]) &  & --- & --- & 13,900$_{-11,200}^{+65,800}$ &  --- & 23,300$_{-17,700}^{+153,500}$ \\
\hline 
$T_e$ RLs &  &  &  &  &  & \\
\ion{He}{i}-$\lambda\lambda$7281/6678 & --- & --- & 8340(330) & 7,800$\pm$800 &  --- & 13,000$_{-6,900}^{+8,400}$ \\
\ion{O}{ii}-\citetalias{peimbert14} & --- & --- & --- & 8,100$_{-400}^{+500}$ &  --- & 7,900$\pm$300 \\
\hline 
$n_e$ RLs &  &  &  &  &  &  \\
\ion{O}{ii}-$\lambda\lambda$4649/61$^a$ & --- & --- & --- & 10,700: &  --- & 11,000: \\
 \hline  
$\mathrm{X^{+i}/H^+}$ CELs &  &  &  &  &  &  \\
N$^+$ ($\times 10^{-5}$) & 1.30 & 0.759 & 3.60 & 1.81$_{-0.17}^{+0.34}$ &  2.31 & 2.82$_{-0.34}^{+0.58}$ \\
O$^+$ ($\times 10^{-5}$) & 5.00 & 2.46 & 27.00 & 14.84$_{-1.65}^{+1.87}$ &  9.10 & 33.49$_{-8.89}^{+16.01}$ \\
O$^{+2}$ ($\times 10^{-4}$) & 2.44 & 2.23 & 1.87 & 0.96$_{-0.08}^{+0.14}$   & 1.01 & 1.44$_{-0.23}^{+0.42}$ \\
Ne$^{+2}$ ($\times 10^{-5}$) & 8.80 & 5.70 & 8.41 & 3.45$_{-0.38}^{+0.61}$ &  2.09 & 2.68$_{-0.47}^{+0.77}$ \\
Ar$^{+2}$ ($\times 10^{-6}$) & --- & 1.44 & 1.59 & 0.96$_{-0.08}^{+0.10}$ &  0.914 & 1.24$_{-0.10}^{+0.14}$ \\
Ar$^{+3}$ ($\times 10^{-8}$) & --- & --- & 2.10 & 0.95$_{-0.94}^{+0.97}$ & --- & --- \\
S$^+$ ($\times 10^{-7}$) & 2.20 & 2.96 & 10.69 & 8.23$_{-2.27}^{+5.52}$ &  9.34 & 9.82$_{-1.89}^{+3.23}$ \\
S$^{+2}$ ($\times 10^{-6}$) & 2.03 & 2.45 & 6.82 & 1.66$_{-0.23}^{+0.35}$ &  1.43 & 2.53$_{-0.49}^{+0.72}$ \\
Cl$^+$ ($\times 10^{-8}$) & --- & --- & 1.76 & --- &  --- & 1.66$_{-0.22}^{+0.30}$ \\
Cl$^{+2}$ ($\times 10^{-8}$) & --- & 3.16 & 10.3 & --- &  --- & 2.80$_{-0.53}^{+1.0}$ \\
Fe$^+$ ($\times 10^{-8}$) & --- & --- & --- & 3.57$_{-1.32}^{+1.61}$ &  --- & 79.23$_{-8.52}^{+11.69}$ \\
Fe$^{+2}$ ($\times 10^{-8}$) & --- & --- & 7.26 & 4.77$_{-2.24}^{+2.72}$ &  --- & 12.98$_{-2.55}^{+3.68}$ \\
\hline 
$\mathrm{X^{+i}/H^+}$ RLs &  &  &  &  &  &  \\
He$^+$ & 0.103 & --- & 0.0969 & 0.089$\pm$0.006 &  --- & 0.062$\pm$0.016 \\
O$^{+2}$-v1 ($\times 10^{-4}$) & --- & --- & 2.82 & 3.38$\pm$0.80 &  --- & 3.61$_{-0.48}^{0.57}$ \\
C$^{+2}$-$\lambda$4267 ($\times 10^{-5}$) & --- & 6.91 & 9.72 &  10.58$_{-4.58}^{+4.80}$ &  --- & 7.40$_{-3.98}^{+3.87}$ \\
N$^{+2}$-$\lambda$4631 ($\times 10^{-5}$) & --- & --- & 6.97 & 9.85$_{-5.33}^{+4.62}$ &  ---  & --- \\

\hline
ADF(O$^{+2}$) & --- & --- & 1.51(0.36) & 3.47$_{-0.89}^{+0.84}$  & --- & 2.52$_{-0.65}^{+0.60}$ \\
\hline 
12+log(X/H) &  &  &  &  &  &  \\
He/H & 11.01 & 10.97 & 10.99 & 10.95$\pm$0.03 &  10.98 & 10.80$_{-0.13}^{+0.08}$ \\
O/H & 8.48 & 8.39(0.10) & 8.66 & 8.39$\pm$0.04 & 8.28(0.19) & 8.68$_{-0.11}^{+0.13}$ \\
N/H-\citetalias{kingsburgh94} & 7.81 & 7.81(0.18) & 8.05 & 7.47$_{-0.06}^{+0.10}$ &  7.76(0.17) & 7.60$_{-0.04}^{+0.08}$ \\
Ne/H-\citetalias{kingsburgh94} & 7.96 & 7.76(0.11) & 8.19 & 7.95$\pm$0.05 &  7.54(0.22) & 7.95$_{-0.13}^{+0.15}$ \\
Ar/H-\citetalias{kingsburgh94} & --- & 6.25(0.11) & 6.37 & 6.25$\pm$0.04 &  6.00(0.18) & 6.36$\pm$0.05 \\
S/H-\citetalias{kingsburgh94} & 6.34 & 6.78(0.14) & 6.83 & 6.41$_{-0.08}^{+0.13}$ &  6.38(0.23) & 6.55$_{-0.08}^{+0.11}$ \\
Cl/H-\citetalias{delgado-inglada14} & --- & 4.73(0.11) & 5.08 & --- &  --- & 4.81$_{-0.07}^{+0.08}$ \\
Fe/H-\citetalias{rodriguez05} & --- & --- & 5.22 & 5.15$_{-0.16}^{+0.14}$ &  --- & 6.12$\pm$0.05 \\
C/H &  & 7.88 & 8.16 & --- &  --- & --- \\
ICFs used &   PTP & KB94 & DI14 &  KB94  &   KB94 & KB94 \\
\hline 
log(N/O)-\citetalias{kingsburgh94} & $-$0.67 & $-$0.58 & $-$0.61 & $-$0.90$\pm$0.07 &  $-$0.52 & $-$1.08$_{-0.08}^{+0.09}$ \\
log(Ne/O)-\citetalias{kingsburgh94} & $-$0.52 & $-$0.64 & $-$0.47 & $-$0.45$\pm$0.02 &  $-$0.74 & $-$0.73$\pm$0.02 \\
log(Ar/O)-\citetalias{kingsburgh94} & --- & --- & $-$2.29 & $-$2.13$\pm$0.04 &   & $-$2.32$_{-0.11}^{+0.09}$ \\
log(S/O)-\citetalias{kingsburgh94} & $-$2.14 & $-$1.62 & $-$1.83 & $-$1.97$_{-0.07}^{+0.10}$ &  $-$1.90 & $-$2.14$\pm$0.07 \\
log(Cl/O)-\citetalias{delgado-inglada14} & --- & $-$3.67 &$-$3.58 & --- &  --- & $-$3.87$\pm$0.09 \\
log(Fe/O)-\citetalias{rodriguez05} & --- & --- & $-$3.44 & $-$3.24$_{-0.17}^{+0.13}$ &  --- & $-$2.56$\pm$0.09 \\
log(C/O) & 0.18 & $-$0.51 & $-$0.50 & --- &  --- & --- \\
 \hline
\multicolumn{7}{l}{$^a$ ":" represents a very uncertain result.} 

\end{tabular}
\end{table*}

\subsection{Physical conditions and ionic abundances from CELs}

Ionic abundances for N$^+$, O$^+$, O$^{+2}$, Ne$^{+2}$, Ar$^{+2}$, Ar$^{+3}$, S$^+$, S$^{+2}$, Cl$^{+}$, Cl$^{+2}$, Fe$^{+}$ and, Fe$^{+2}$, were determined for the ESO VLT UVES and SALT HRS data, from the observed CELs using the task \textit{get.IonAbundance} from \textsc{PyNeb}. For this we employed the dereddened line intensities listed in Table \ref{tab:intensities} as measured for each data set, and  the adopted physical conditions for each nebular zone as presented in Table \ref{tab:chemistry}.  Since the temperatures derived from [\ion{N}{ii}], [\ion{O}{iii}] and [\ion{Ar}{iii}] line ratios are similar, only one electron temperature was used for the whole nebula, which is the mean value between the temperatures mentioned (the [\ion{Ar}{iii}] temperature is only available for SALT data). We have verified that this adoption for the temperature value produces equal ionic abundances by using the nebular and auroral lines of the ions. For the density, from data from ESO UVES only a single zone given by [\ion{S}{ii}] density was assumed, while for SALT data the [\ion{S}{ii}] density value was used for the low-ionization zone and the [\ion{Cl}{iii}] density for the more ionized zone. Density values from [\ion{Ar}{iv}] and [\ion{Fe}{iii}] were not considered due to their large  uncertainty.

Ionic abundances are given in Table  \ref{tab:chemistry}. In the same table we included the temperatures, densities, ionic and total abundances presented by the different authors in the period from 1992 to 2021. These data have been taken directly from the publications, without any modification.

In all cases,   [\ion{O}{iii}] electron temperatures  are in the range from 10,100 to 11,500 K and [\ion{N}{ii}] temperatures are in the  10,600 to 12,900 K range, except for the temperatures  derived by \citet{otsuka17} which are about 2,000 K lower in both cases. The same occurs for the temperature from [\ion{Ar}{iii}]. Most probably this is due to \citet{otsuka17} used a different set of atomic parameters for both emission-line analyses (CELs and RLs). As they mentioned,  effective recombination coefficients,
transition probabilities, and effective collision strengths listed
in \citet{otsuka10} were used, which are different from the parameters used by other authors. 

\subsection{Physical conditions and ionic abundances from RLs}

In this work electron density from \ion{O}{ii} recombination lines can be derived from the  ratio $\lambda\lambda$4649/4661. The derived values are similar to the values from CELs in both, ESO VLT UVES and SALT HRS observations, however both values present very high uncertainties and thus the density from CELs was adopted for the RLs abundance calculations. 

Temperatures from RLs can be derived from \ion{He}{i} and \ion{O}{ii} lines. We followed the methodology from \citet{zhang05} for \ion{He}{i} temperature using the line ratio $\lambda \lambda$7281/6678.  Equation 1 by \citet{zhang05} and the coefficients $a_i$, $b_i$ and $c_i$ for these lines determined by \citet{benjamin99}, for a $n_e = 10^4$ cm$^{-3}$, which is adequate to the adopted density for ORLs, were used. To derive temperature from \ion{O}{ii} lines we followed the methodology proposed by \citet{peimbert14}, which requires the intensities of \ion{O}{ii} V1 multiplet and of [\ion{O}{iii}] $\lambda\lambda$4959,5007 lines. The derived values with their uncertainties are listed in Table \ref{tab:chemistry}. As expected it is found that both temperatures are lower than the temperatures derived from CELs.

$T_e$(\ion{He}{i}) was used to determine ionic abundances of He$^{+}$ while $T_e$(\ion{O}{ii}) was used for all other RLs ionic abundances.  Such abundances  were computed using \textsc{PyNeb} routine 
{\it get.IonAbundance} by using the  temperatures and densities mentioned above. The computed values for He$^+$, O$^{+2}$, C$^{+2}$ and N$^{+2}$ abundances are listed in Table \ref{tab:chemistry} where the emission lines used for these calculations are marked. The comparison of O$^{+2}$ abundances derived from CELs and from RLs allows the determination of the Abundance Discrepancy Factor, ADF, defined as the ratio between the abundance from RLs and the abundance from CELs. The ADFs(O$^{+2}$) derived from our data are 3.47 for the ESO VLT UVES data and 2.52 for the SALT data and they are presented in Table \ref{tab:chemistry}.

\subsection{The temporal behaviour of ionic abundances derived from CELs}

By analysing the temporal evolution of ionic abundances presented in Table \ref{tab:chemistry} interesting systematic behaviours are found. In this table it is noticed that O$^{+2}$/H$^+$ ratio has been decreasing from a value larger than $2.0 \times 10^{-4}$ in 1992 to $1.44 \times 10^{-4}$ in 2021. The same happens with ionic abundances of other highly ionized species like Ne$^{+2}$, Ar$^{+3}$, and Ar$^{+2}$. On the other hand, the relative abundance of low-ionized species such as O$^+$, S$^+$, N$^+$, and Fe$^{+2}$ are increasing with time. All this indicates that the  highly-ionized species are  recombining in response to the fact that the central star has been cooling down with time.

The fast nebular recombination  of highly ionized species is a direct consequence of the high electron density and the fast  evolution of the central star. It is known that for a nebula of about 6,000 cm$^{-3}$ the recombination time of H$^+$ is less than 20 yr once the ionizing photons flux is closed, and the recombination time of O$^{+2}$ is faster due to the larger recombination coefficient (see e.g., \citealp{osterbrock06}, Appendix 5, Table A5). The electron density appears to be larger in some zones of  Hen\,3-1357, therefore a fast recombination of the highly ionized species is expected as the star cools down.

\section{Derived total abundances. Comparison with previous determinations}

Total abundances were determined from the ionic abundances and using some Ionization Correction Factors (ICFs) commonly used in the literature, to correct for the unseen ions. The references for the ICFs employed, their expressions  and used values are presented in Appendix A. In Table \ref{tab:chemistry} we present the abundances derived from our data indicating the ICFs used.  Also the abundances derived by the different authors are included. Abundances are not expected to vary with time, therefore values derived by the different authors should be similar.

- The values determined for helium abundance between 1990 and 2011 in the different papers are $\rm{12+log(He/H) = 10.97 - 10.98}$. Thus the He/H ratio in Hen\,3-1357 appears 0.05 dex higher than solar (a value of 10.93 was determined by \citealp{asplund09}), and slightly lower than the average value found in  non-Type I PNe \citep{kingsburgh94}\footnote{Non-Type I PNe are those belonging to the galactic disc, with N/O abundance ratio $\leq$ 0.5 by number. The definition of PN Types was proposed by \citet{peimbert78}}. 
This value decreases in the observations of SALT 2021 ($\rm{12+log(He/H) = 10.80}$), because He$^+$ is recombining into He$^0$ which is not considered when determining the total He abundance.

- The  derived 12+log(O/H) values are in the range from 8.4 to  8.6. Therefore the O abundance appears slightly sub solar \citep[solar 12+log(O/H) is 8.69 according to][]{asplund09}.  The value presented by \citet{otsuka17} is slightly larger than the other values due to the lower electron temperatures derived by these authors which induce very large values for O$^+$/H$^+$, N$^+$/H$^+$, S$^+$/H$^+$ and other ions. A similar fact occurs with the SALT-HRS data, where also a low temperature is found, although with very large errors.

- Regarding nitrogen, 12+log(N/H) is around 7.7 in all cases (except in \citealt{otsuka17} where is larger) which is a sub solar value \citep[the solar value is 7.83 as derived by][]{asplund09}. 

- Most 12+log(S/H) values are around 6.5, a sub solar value \citep[][reported 7.12]{asplund09}. 

- 12+log(Ne/H) shows values around 7.95, which is very similar to the solar abundance \citep[a value of 7.93 is given by][]{asplund09}. A larger value of 8.19 is found in \citet{otsuka17}. Possible Ne is enhanced by the production of $^{22}$Ne during stellar nucleosynthesis.

- 12+log(Ar/H) values derived by us are around  6.2, which is a sub solar value \citep[][reported 6.40 for the Sun]{asplund09}. 

- 12+log(Cl/H) values are between 4.7 and 4.8. A higher value of 5.08 is reported in \citet{otsuka17}. In any case, Cl/H abundance appears highly sub solar, compared with the value of 5.50 reported by \citet{asplund09}.

- In the case of Fe, the derived values of 12+log(Fe/H) go from 5.2 to 6.1 and are largely sub solar, more than a hundred times lower than the solar value that according to \citet{asplund09} is 7.50. There seems to be a large amount  of dust in this nebula. \citet{otsuka17} claimed that the largely depleted Fe/H ratio suggests that over 99\% of the Fe atoms in the nebula would be locked within silicate grains.

\section{Discussion}

The N, Ne, Ar and Cl abundances, relative to O, presented in Table \ref{tab:chemistry} show that the Stingray nebula seems to be a normal non-Type I planetary nebula with sub solar abundances. 

 Nebular abundances can be compared to the predictions of stellar nucleosynthesis models for low-intermediate mass stars at the end of the AGB phase, computed for different masses and different metallicities, to estimate the initial mass of the progenitor star. For instance models  by \citet{karakas10} and \citet{ventura17} (the latter include dust formation) are suitable for this purpose.  We found that the chemical abundances of Hen\,3-1357 are in agreement with the abundances in  \citet{ventura17} models with metallicity Z=0.008. Such models have initially $\rm{He/H} = 10.95$, $\rm{C/H} = 8.05$, $\rm{N/H} = 7.806$, $\rm{O/H} = 8.56$ and $\rm{Ne/H} = 7.806$, in units of 12+log(X/H), and evolve towards larger amounts of C and N for masses increasing from 1 to 8 M$_{\odot}$ (see Fig. 1 by \citealp{ventura17}). Considering the N/H and O/H abundances ratios in Hen\,3-1357, we can conclude that the star SAO\,244567 had an initial mass between 1.0 to 1.5  M$_{\odot}$ because the nebular N is not enhanced as it should be for stars with initial masses equal or larger than 2 M$_{\odot}$. A similar result was obtained by \citet{otsuka17} analysing \citet{karakas10} models. 

However the fast variations of the central star, which in about 20 years evolved from a post-AGB B1 supergiant to a young planetary nebula central star with an effective temperature of about 60,000 K around 2002, and afterwards started cooling down up to 41,000K in 2011 \citep{arkhipova13} do not correspond to the normal evolution of such a single low-mass star in advanced stage of evolution. 

The subsequent fall in the nebular excitation degree, which we have demonstrated in this work, shows that the stellar temperature is still decreasing. By using the procedure described by \citet{arkhipova13} based on an empirical relation determined by \citet{kaler78}, which uses the [\ion{O}{iii}]5007/H$\beta$ line ratio,  we estimated the present effective temperature of the central star to be less than about 40,000 K which confirms that SAO\,244567 is still cooling down. 

 A Late Thermal Pulse (LTP) has been suggested to explain these changes. If such a hypothesis is confirmed,  the star should continue cooling down all the way towards the AGB phase.

Following the fast evolution of the central star, the nebula has shown rapid changes. The  ionization degree of its heavy elements, which increased fast in the 1970 to 1990 period is now decreasing and there is much evidence that such elements are presently recombining. 

\section{Conclusions}

From high resolution spectra obtained in 2009 with the ESO VLT UVES spectrograph, and in 2021 with the SALT HRS spectrograph, physical conditions and chemical abundances have been obtained for the extraordinary and fast evolving planetary nebula Hen\,3-1357, the Stingray Nebula. Our line intensities, physical conditions and chemistry have been compared with values from the literature, derived from 1990 to 2021. 
We have confirmed that the nebula is presently recombining as a result of the fast cooling of the central star whose effective temperature has decreased from about 60,000 K in 2002 to less than 40,000 K in 2021.

The chemical abundances have been derived from CELs for a number of elements. Also chemical abundances have been derived from RLs and the derived ADF(O$^{+2}$) has a value of 3.5 from ESO VLT UVES data and 2.6 from SALT HRS data. This value for the ADF(O$^{+2}$) is within the usual values derived for PNe  (see \citealt{mcnabb13} for a compilation of ADF values in a large sample of PNe). A slightly lower ADF(O$^{+2}$) value of 1.51 was derived by \citet{otsuka17} for Hen\,3-1357.

Our abundance values are in general agreement with values from the literature, and show that Hen\,3-1357 have sub solar O, N, S, and Ar abundances, corresponding to a central star of initial mass similar or lower  than 1.5 M$_{\odot}$. 

The fast evolution of the central star of such a mass is unexpected and it has important consequences in the nebular behaviour which is fast recombining. 
If the central star experienced a LTP, as claimed by \citet{reindl14,reindl17} and \citet{lawlor21}, it should keep cooling down in its way towards the AGB phase. Both, star and nebula, require a close follow up in order to get deeper insight of this phenomenon and to improve our understanding of the low-mass stellar evolution.

\section*{Acknowledgements}

 Some of the observations reported in this paper were obtained with the Southern African Large Telescope (SALT). We are thankful  to Prof. Patricia Whitelock for her help in getting the SALT time.
F. Ruiz-Escobedo acknowledge scholarship from CONACyT-M\'exico. M. Parthasarathy is thankful to Prof. Keivan Stassun, and dean, chairman and faculty of the department of Physics and Astronomy, Vanderbilt University, for their kind hospitality and support. This work received financial support from PAPIIT-UNAM grant IN105020.  

\section*{Data Availability}
The data underlying this article will
be shared on reasonable request to the corresponding author.







\appendix
\section{Ionization Correction Factors}
\label{section:icfs_exp}

Expressions and ICFs used for the total abundances calculation are listed next. 

\begin{itemize}

\item $\rm{ \frac{He}{H} = \frac{He^{+}}{H^{+}} }$. 

\item $\rm{ \frac{O}{H} = \frac{O^+ + O^{+2}}{H^+}}$, $\rm{ICF(ESO-09) = ICF(SALT-21) = 1.00}$.


\item $\rm{ \frac{N}{H} = ICF(N) \times \frac{N^+}{H^+}}$. $\rm{ICF(N) = \frac{O}{O^+}}$ \citep{kingsburgh94}. $\rm{ICF(ESO-09) = 1.65, \, ICF(SALT-21) = 1.44}$.

\item $\rm{ \frac{Ar}{H} = ICF(Ar) \times \frac{Ar^{+2} + Ar^{+3} + Ar^{+4}}{H^{+}} }$, $\rm{ICF(Ar) = \frac{1}{1-N^+/N} }$. If only Ar$^{+2}$ is detectable, $\rm{ \frac{Ar}{H} = 1.87 \times \frac{Ar^{+2}}{H^{+}} }$ \citep{kingsburgh94}. $\rm{ICF(ESO-09)=2.54, \, ICF(SALT-21)=1.87}$.

\item $\rm{ \frac{Ne}{H} =  ICF(Ne) \times \frac{Ne^{+2}}{H^{+}} }$. $\rm{ICF(Ne) = \frac{O}{O^{+2}} }$ \citep{kingsburgh94}. $\rm{ICF(ESO-09)=2.53, \, ICF(SALT-21)=2.19}$.


\item $\rm{ \frac{S}{H} = ICF(S) \times \frac{S^{+} + S^{+2}}{H^{+}} }$. $\rm{ICF(S) = \left[ 1 - \left( 1 - \frac{O^+}{O} \right)^3 \right]^{-1/3}}$. \citep{kingsburgh94}. 
$\rm{ICF(ESO-09) = 1.02, \, ICF(SALT-21) = 1.01}$.


\item $\rm{ \frac{Cl}{O} = ICF(Cl) \times \frac{Cl^{+} + Cl^{+2}}{O^+}}$. $\rm{ICF(Cl) = 1}$ \citep{delgado-inglada14}. $\rm{ICF(ESO-09) = --, \, ICF(SALT-21) = 1.00}$.

\item $\rm{ \frac{Fe}{O} = \frac{Fe^{+} + Fe^{+2}}{O^{+}}}$, \citep{rodriguez05}. 
$\rm{ICF(ESO-09) = ICF(SALT-21) = 1.00}$.

\end{itemize}



\section{Atomic data}

\begin{table*}
\centering
	\caption{\bf Atomic parameters used in \textsc{PyNeb} calculations}
\begin{tabular}{lcc} \hline
Ion & Transition probabilities & Collisional strenghts \\
\hline
N$^+$ & \citet{froese04}  & \citet{tayal11}\\
O$^+$ & \citet{froese04} & \citet{kisielius09}\\
O$^{+2}$ & \citet{froese04} & \citet{storey14} \\
         & \citet{storey00} &  \\
Ne$^{+2}$ & \citet{galavis97} & \citet{mclaughlin00} \\
S$^+$ & \citet{podobedova09} & \citet{tayal10} \\
S$^{+2}$ & \citet{podobedova09} & \citet{tayal99} \\
Cl$^{+}$ & \citet{mendoza83} &  \citet{tayal04} \\ 
Cl$^{+2}$ & \citet{mendoza83} & \citet{butler89} \\
Ar$^{+2}$ & \citet{mendoza83} & \citet{galavis95} \\
         & \citet{kaufman86} &           \\
Ar$^{+3}$ & \citet{mendoza82}  & \citet{ramsbottom97} \\
          & \citet{kaufman86}  & \\
Fe$^{+}$ &  \citet{bautista15} & \citet{bautista15}   \\
Fe$^{+2}$ & \citet{quinet96}    & \citet{zhang96}  \\
          & \citet{johansson00} &           \\
\hline

Ion & \multicolumn{2}{c}{Effective recombination coefficients} \\
\hline
H$^+$ & \multicolumn{2}{c}{\citet{storey95}} \\
He$^+$ & \multicolumn{2}{c}{\citet{porter12,porter13}} \\
N$^{+2}$ & \multicolumn{2}{c}{\citet{fang11}}  \\
O$^{+2}$ & \multicolumn{2}{c}{\citet{storey17} }  \\
C$^{+2}$ & \multicolumn{2}{c}{\citet{pequignot91}}  \\
\hline
\end{tabular}
\label{tab:atomic-parameters}
\end{table*}


\bsp	
\label{lastpage}
\end{document}